**Realizing a High Magnetic Moment in Gd/Cr/FeCo: The Role of the Rare Earth**

C. Ward[1]*, G. Scheunert[1], W.R. Hendren[1], R. Hardeman[2], M.A. Gubbins[2] and R. M. Bowman[1#]

[1]Centre for Nanostructured Media, School of Mathematics and Physics, Queen's University Belfast, Belfast BT7 1NN, UK

[2]Seagate Technology, 1 Disc Drive, Springtown, Northern Ireland, BT48 0BF, UK

**Abstract**

The search for materials or systems exhibiting a high magnetic saturation has been of longstanding importance. It has been suggested that increased saturation could be achieved by coupling a transition metal via a spacer to a rare earth. We report Gd/Cr/Fe$_{70}$Co$_{30}$ multilayer stacks and find reduced yet modulating magnetic moment as a function of Cr thickness. Through a micro structural analysis the lowered moment is indicated by the nucleation of the ultrathin Gd films into an fcc phase. We discuss the possible solution in terms of quasi-perfect lattice match seed material to promote growth of hcp Gd.

Corresponding authors: * cward19@qub.ac.uk  # r.m.bowman@qub.ac.uk

Going back over decades there has been a long-standing desire to explore routes that would lead to a saturation magnetization in excess of the limit established by the Slater-Pauling rule of around 2.45T in the Iron Cobalt alloy at room temperature[1]. This has lead to wide-ranging studies in a vast array of compounds and superlattices[2,3]. The common failing of most potential systems is that volume expansion effects are ignored, overlooked or underestimated and then leads to spurious claims for enhanced saturation magnetization e.g. the Iron nitrides being a good example[4]. In tandem, effort has been made to utilize the high magnetic moment per atom, in the lanthanides e.g. Gd ($7.6\mu_B$) and Dy ($10.6\mu_B$)[5]. These Rare Earth (RE) elements possess low Curie temperatures and so they present a further and particular challenge for many applications. Notwithstanding this the REs have been extensively studied over the last two decades by examining the effect of multilayers (Dy/Er[6] and LuY/Dy[7]), and for example Gd/W systems with regard to increasing the Curie temperature and enhanced remnant magnetisation[8]. At the same time there is the observation of an improved interfacial coupling in Gd/Fe multilayers[9] but with the recognition of antiferromagnetic (AF) coupling reducing the overall magnetic moment, as is the case in Gd/Y multilayers[10].

More recently there has been the suggestion to employ the use of a metal (M) spacer layer such as Chromium to mediate the coupling[11]. This was based upon the recognition that while Fe couples AF to Cr[12], Cr couples AF to Gd[13]. This then raised the possibility of an overall ferromagnetic (FM) coupling between the 3$d$ transition magnetic metal such as Fe and the RE. This was supported by density functional theory calculations that informed the synthesis of a series of Gd/Cr/Fe multilayer systems[11]. Examination of the multilayers with X-ray magnetic circular dichroism (XMCD) measurements identified coupling as evidenced by changes in the $T_c$ of the multilayer system. However, there has been no further discussion of the actual *functional* magnetic response of such a system. On this basis of the above it was further postulated that greater potential lay in the substitution of the Fe for the higher saturation of the $Fe_{70}Co_{30}$ system[14].

If there is indeed a coupling between the RE to the FM then it is timely to establish the extent of the magnetic functional response. Therefore in this letter we report the results of a careful study of the coupled $Gd/Cr/Fe_{70}Co_{30}$ system, with a varying Cr thickness, and present the magnetic behavior. While we see evidence for coupling we identify a more fundamental issue that challenges the utilization of REs, namely the longstanding but infrequently identified issue of a paramagnetic fcc phase and its role in ultrathin layers of RE like Gd.

Experimentally, we commenced the work with the preparation of samples via UHV DC magnetron sputtering in a Lesker co-deposition system onto 75 and 150mm $Si/SiO_2$ wafers at room temperature and elevated temperatures (350 $^{o}$C). For increased uniformity the substrate was rotated at 20RPM. Base pressure was <5 x$10^{-9}$ mbar, pressure of the process gas Ar was 4 x$10^{-3}$ mbar. Deposition rates were calibrated via x-ray diffraction (XRD) and transmission electron microscopy (TEM). Rates were also monitored real-time with a quartz crystal. During elevated temperature depositions the sample holder and wafer were allowed significant time to heat up to outgas, with monitoring by residual gas analyzer, to allay any fears of oxidation from these sources. All samples were deposited with a seed layer of 5nm Ta. At the outset we validated the formation of our Chromium layers by demonstrating RKKY coupling in $Fe(4nm)/Cr(x)/Fe(4nm)$, where x ranges from 0.2nm to 2nm. The $Si/SiO_2/Ta/Gd/Cr/Fe_{70}Co_{30}$ samples described are capped with 2nm of Ta as oxidation of RE elements is always of significant concern as even a small contamination can degrade the magnetic properties[15].

Characterisation of the samples in-plane magnetic properties was done using a Quantum Design MPMS XL SQUID magnetometer with applied fields of up to 5T. The instrument was calibrated with other metrology tools and NIST standard samples, the magnetization data presented was also carefully corrected for the diamagnetic response of the wafer and contribution from the sample holder.

For this study we prepared series of multilayer samples based on $Gd(6nm)/Cr (nm)/Fe_{70}Co_{30} (4.23nm)$ where the thickness of Cr varied from 0.2nm to 2.4nm. The Gd is prepared with significant care[16]. In Figure 1(a) we show the dependence of the system's saturation magnetization ($M_S$) with temperature. It is evident that there is considerable variation of temperature response of $M_S$ with the changing thickness of Cr. In Figure 1(b) we show Ms values at 4K and there is modulation of the magnetization akin to RKKY coupling[13]. The largest values of $M_S$ occur for Cr thickness of 0.8nm and 1.0 nm and the lowest at 1.4nm and 1.8nm. For the RE/M/FM system it is postulated that the coupling sign oscillates with FM/RE separation[11] where an odd number of monolayers of Cr (0.145nm) gives anti-ferromagnetic coupling and even number ferromagnetic. While we do see a modulation of saturating magnetization with varying spacer distance there is not an exact corroboration with the model. We would attribute this to minor fluctuation in our 0.7 Å/s Cr deposition rate. However, attention is drawn to the observation that for all of our samples the total magnetization is significantly suppressed as compared to the magnetization of our typical 50 nm Gd of 2.6T (at 4K)[16] and to even that of the 4.23 nm $Fe_{70}Co_{30}$ at 2.3T. In the Figure 1 we also include a $M_S$ (T) curve for a single 6 nm Gd film for comparison.

From this position we sought to investigate the suppression of $M_S$ in these ultrathin Gd layers by the preparation of further multilayer systems with increased Gd thickness. Data for these are shown in Figure 2. There is now two distinct changes for the Gd/Cr/$Fe_{70}Co_{30}$ stacks with increasing Gd thicknesses; the saturation magnetization of the system has increased significantly and there is now a temperature dependence more like that of a typical Gd single layer.

The trend in the $M_S$ for Gd thickness of 6, 12, 25 and 50 nm clearly points to an improving quality of the Gd as reflected in the magnetization. Whilst tempting at this stage to consider arguments around improved coupling it is necessary to revisit the microstructure of the thinnest Gd layers. There have been several reports of Gd favoring to nucleate at a seed boundary in a paramagnetic fcc phase[17,18] particularly in thin films[19,20]. To examine this a series of θ-2θ XRD measurements were taken of the films and multilayers and these are shown in Figure 3. For the thicker Gd films we observe the dominance of the desired hexagonal close-packed (hcp) structure indicated by the (002) peak with a = (3.60±0.02) Å and c = (5.77±0.02) Å in good agreement with single-crystal bulk Gd[5]. However, for the multilayers and most noticeable in the single layer of 6nm Gd is that while there is the occurrence of the expected hcp reflections there is also the obvious presence of an fcc contribution, indicated by the (111) peak. This fcc phase appears with a = (5.31±0.02) Å and is in agreement with that found previously[19] yet a slight peak shift towards larger angles corresponding to a 0.03 Å decrease which would indicate some strain in the layers. Previously, we demonstrated through grazing incidence XRD scans, that this phase was likely to exist at the seed layer interface[16] as Ta nucleates bcc[21] and hence provides an imperfect structural dictate[16], and was most likely due to stacking faults[22,23]. As the Gd thickness increases we see a declining influence of the fcc phase and the ferromagnetic hcp becomes dominant. This XRD data gives a clear indication of the likely parasitic role that the fcc phase plays in determining the magnetics of the whole multilayer system. We note the extreme difficult of examining the multilayers by electron microscopy as sample preparation leads to the formation of not only amorphous surfaces but also $Gd_2O_3$ in the thinnest specimens.

To examine the hypothesis that the Gd nucleates fcc we now present data taken from a 4nm Gd film in Figure 5. We observe several features; the film is almost exclusively fcc with no hcp peak evident, the fcc reflection is also very broad and comprises extremely small grains of the order of 5nm. In Figure 5(b) and (c) magnetic data show that the sample has reduced saturation and the low field measurements indicate a suppressed Curie temperature of around 225K. Most significantly we now note that the phase is not paramagnetic; the low field measurements (100 Oe) show that it exhibits a remnant field up to 125K. The low coercivity of this fcc phase will clearly lead to it playing a disproportionate influence in our multilayers comprising 6nm Gd.

The observation that this fcc phase is not paramagnetic but has ferromagnetic contribution could allow effective thickness models to be developed. At one limit, if the original saturation magnetization values (Figure 1) are recalculated by discarding an interfacial region of fcc Gd of 5 nm thickness, illustrated in Figure 4, we see revised values of $M_S$ more akin to what we would expect for single phase films. Assuming a worst-case scenario of say 10nm fcc strata would yield a maximum (4 K) $M_S$ of 2.7 T, slightly greater than what expected for a maximum obtainable in Gd. The answer for the thickness of fcc lies somewhere in between.

External corroboration of the observation that a parasitic fcc phase is present and so degrades functional magnetic response occurs if we revisit the microstructure of the originally proposed Gd/Cr/Fe where a more recent report clearly indicated that at the thickness (4.34nm) of Gd employed there is a similarly large fraction of the Gd fcc phase present[24], though it is not indexed. Consequently, taking our data here along with the absence of functional data in other work[11,24] we believe that "huge magnetization" of such RE/M/FM multilayers cannot be practically realized until this paramagnetic portion can be removed from the Gd or other RE film in such multilayers.

The most obvious route to developing a purely hcp RE film is by consideration of suitable epitaxy. Most of the rare earth metals share similar structure and exhibit a hexagonal close-packed crystal structure at room temperature, such as Gd, Dy, Ho, Tb and Er. Lattice parameters are in the range of a = 3.56-3.63Å and c = 5.59 − 5.78Å, where Gd has the largest unit cell[5]. Previous reports on thin-film growth modes indicate very similar nucleation for all of these elements. Independent of the thin-film deposition method such as sputtering, evaporation or MBE, they exhibit Stranski-Krastanov growth, i.e. a smooth initial layer of 5-10nm thickness is followed by island formation. After the first reports on the occurrence of what was termed a metastable fcc phase for thin films of Gd[19] a similar fcc phase was found in Dy and Ho[20]. It can be surmised that if RE thin films (< 5nm) do prefer to initially order fcc and only turn hcp with further nucleation and thickness then the only viable route to establishing the efficacy of the coupling proposed is to form a structure with an epitaxial Gd layer by preparing it on a material with perfect lattice match.

In summary we have shown that the inclusion of thin layers of Cr can have a significant impact on the saturating magnetisation of Gd and $Fe_{70}Co_{30}$ layers. We have presented evidence to show that this increase is probably due to an effective ferromagnetic coupling between the two layers moderated by the Cr spacer. However, analysis of the magnetics and structure for the thin Gd shows that it is still significantly influenced by the occurrence of an fcc phase contribution. This then illustrates why no conclusive evidence of enhanced magnetisation as been demonstrated in such a RE/M/FM system. However our results point a clear path towards a practicable realization to test the hypothesis.

## Acknowledgements

We thank Seagate Technology (Ireland) for their financial support to establish ANSIN (www.ansin.eu)

<u>Figure Captions</u>

Figure 1. (a) In plane saturation magnetisation (H = 50kOe) of Gd (6nm)/Cr (*x*nm)/Fe$_{70}$Co$_{30}$ (4.23nm) showing modulation. Included are Gd (6nm) and Fe$_{70}$Co$_{30}$ (4.23nm) for reference purposes. (b) Maximum saturating magnetisation taken at 4K (H=50kOe) with respect to Cr interlayer thickness highlighting an oscillatory nature of the coupling.

Figure 2. In plane saturation magnetization (H=50kOe) of Gd(50nm)/Cr(1.3nm)/ Fe$_{70}$Co$_{30}$ (4.23nm) and Gd(25nm)/Cr(1.3nm)/ Fe$_{70}$Co$_{30}$ (4.23nm) showing influence of thickness of Gd layer.

Figure 3. XRD spectra (Θ-2Θ) of samples of **(a)** Gd(50nm)/Cr(1.3nm)/ Fe$_{70}$Co$_{30}$(4.23nm), **(b)** Gd(25nm)/Cr(1.3nm)/Fe$_{70}$Co$_{30}$(4.23nm),**(c)** Gd(12nm)/Cr(1.3nm)/Fe$_{70}$Co$_{30}$(4.23nm), **(d)** Gd(6nm)/Cr(1.3nm)/Fe$_{70}$Co$_{30}$(4.23nm) and **(e)** Gd(6nm) Reference, sputtered on a Si/SiO$_2$ wafer at 350°C, the presence of significant proportions of fcc Gd for thinner layers should be noted and also the evolution of hcp Gd with respect to the fcc phase for thicker depositions.(As a guide the 29.1° and 30.95° 2Θ angles are highlighted for clarity)

Figure 4. Attempted correction of Figure 2 data of Gd (50nm)/Cr (1.3nm)/ Fe$_{70}$Co$_{30}$ (4.23nm). Correction assumes no contribution from selected fcc strata and shows effect on magnetic data when the strata is 5 and 10nm.

Figure 5. (a) XRD spectra (Θ-2Θ) of samples of Gd (4nm) sputtered on SiO$_2$ showing a purely fcc composition. (b) In plane saturation magnetization (H= 10KOe) and remnant field of Gd (4nm), (c) In plane ZFC (--->) - FC (<---) moments (H= 100Oe) of Gd (4nm) layer showing magnetic transitions.

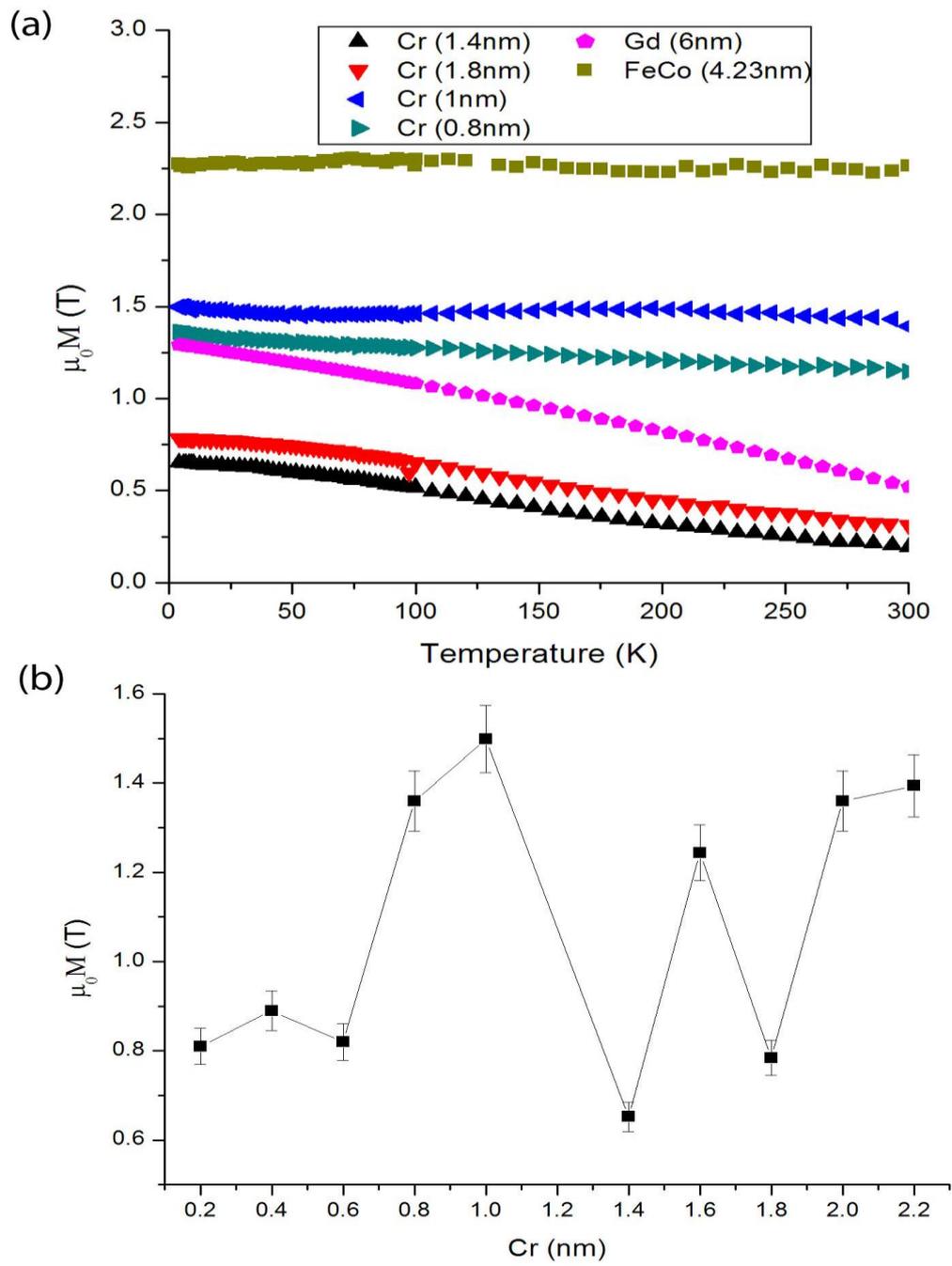

Figure 1

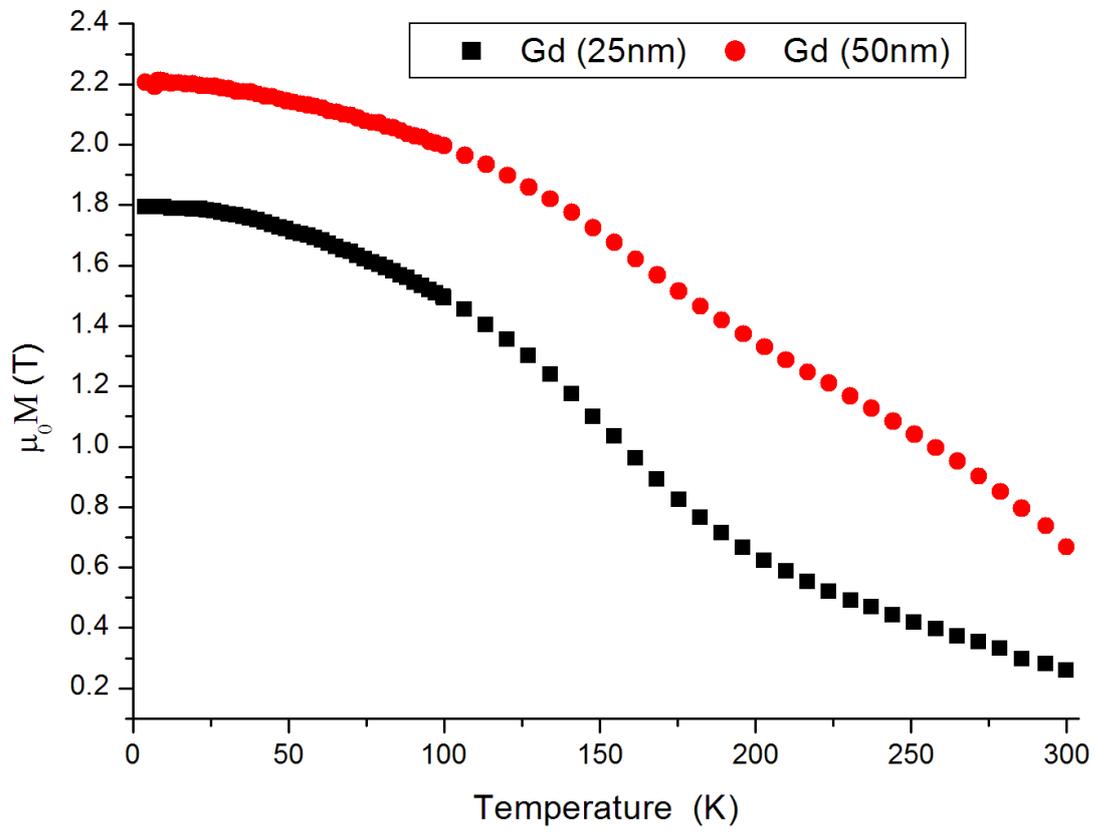

Figure 2

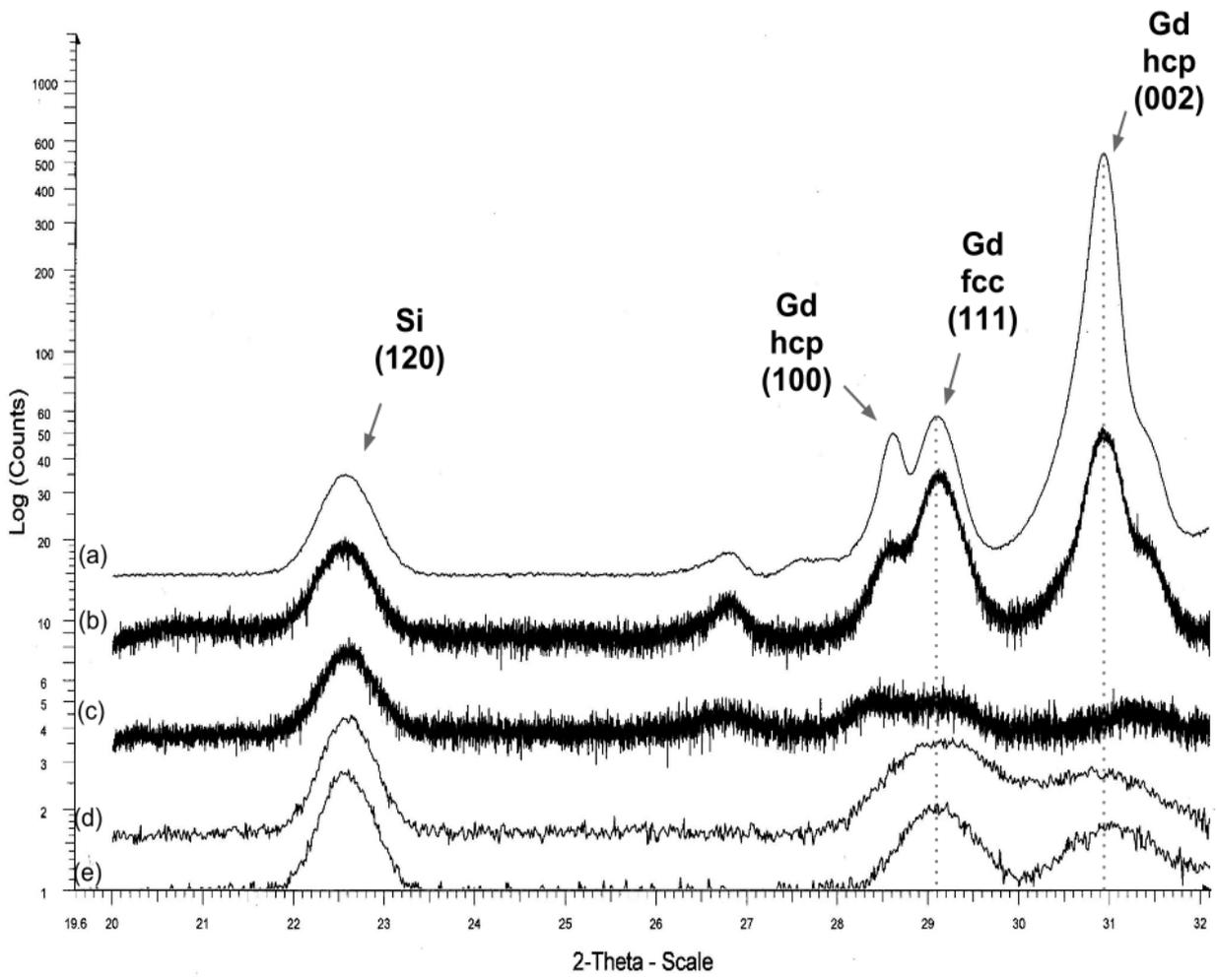

Figure 3

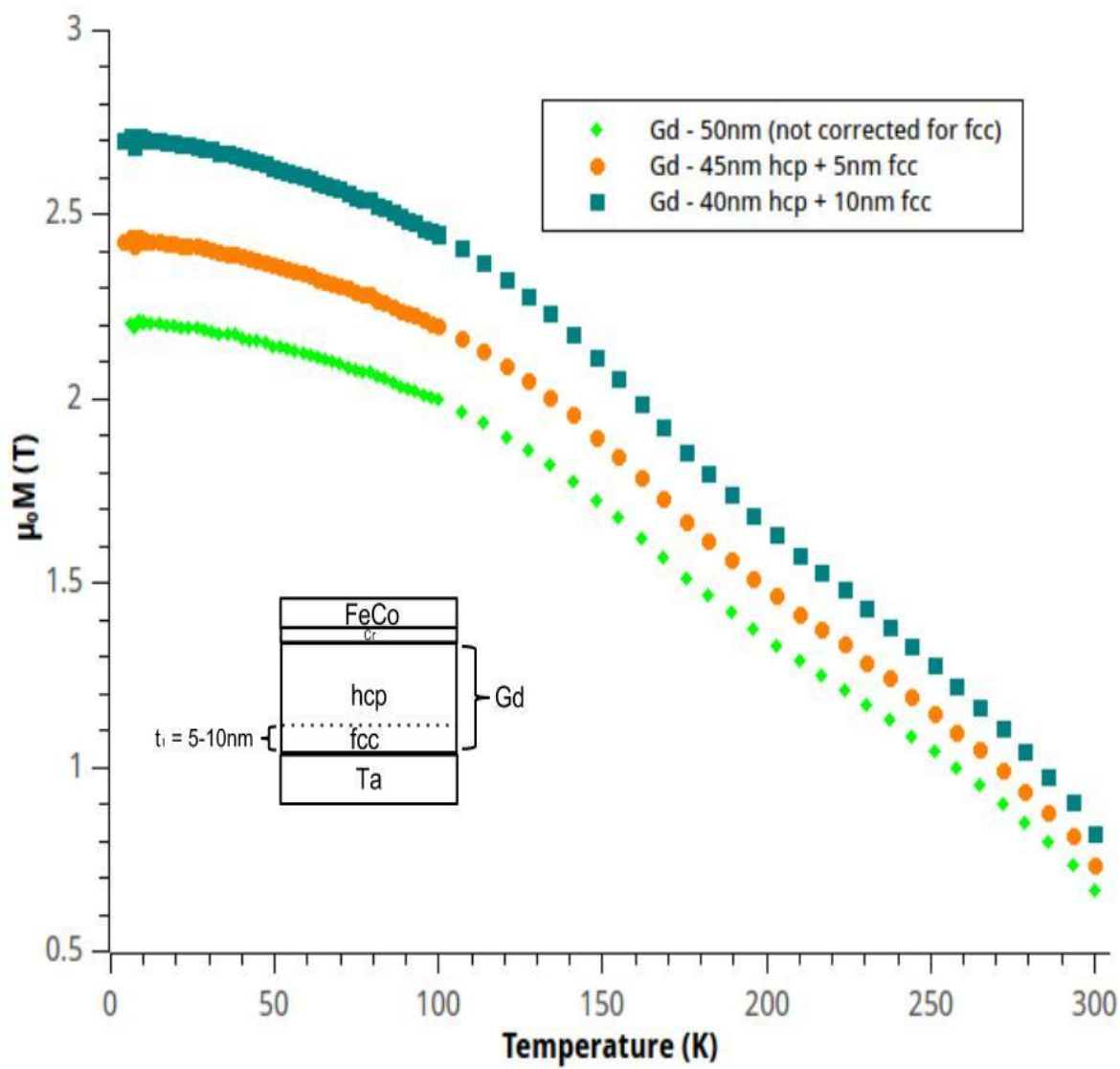

Figure 4

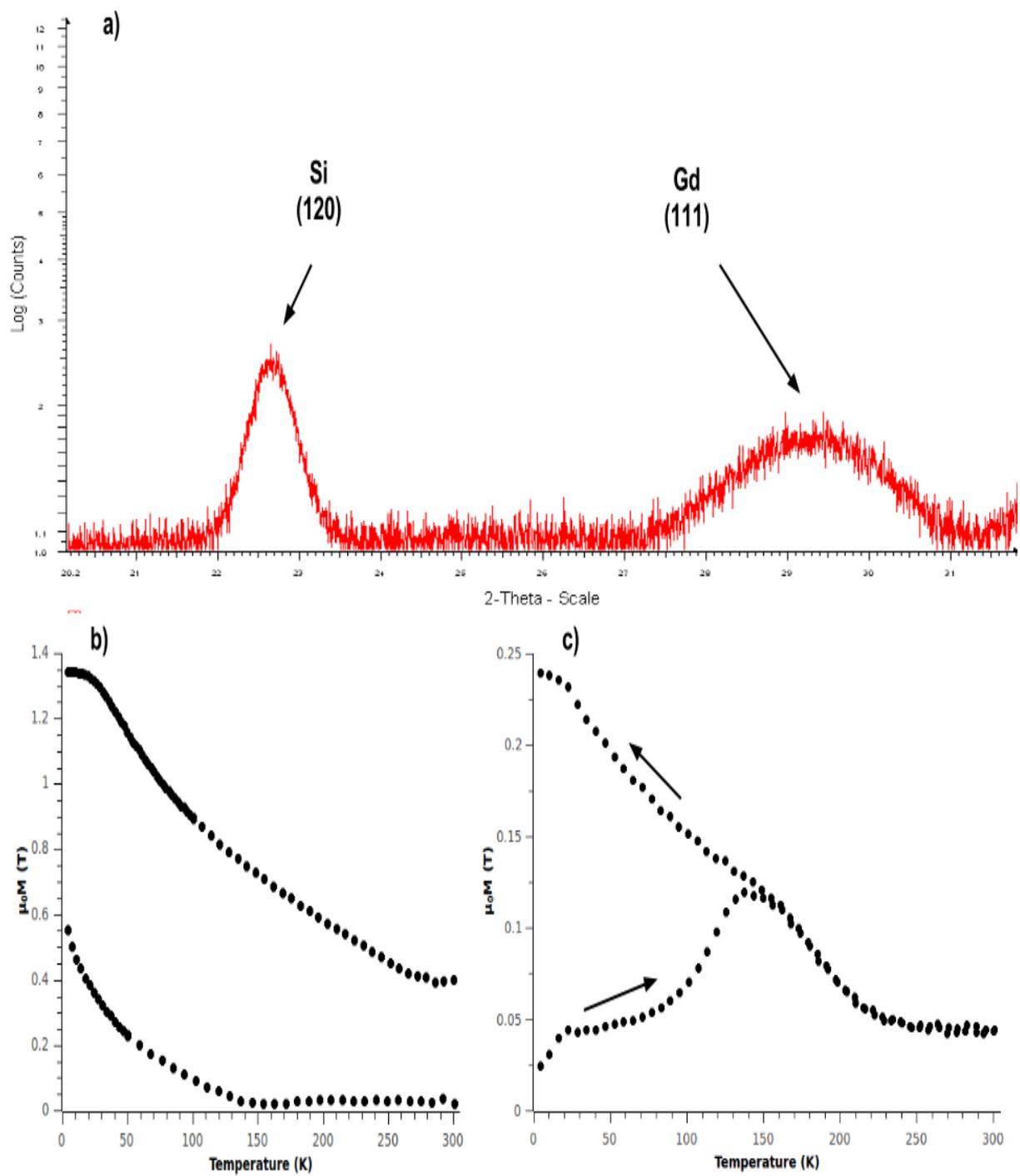

Figure 5